# Reversible redox reactions in an epitaxially stabilized SrCoO$_x$ oxygen sponge

**Hyoungjeen Jeen[1], Woo Seok Choi[1], Michael D. Biegalski[2], Chad M. Folkman[3], I-Cheng Tung[4,5], Dillon D. Fong[3], John W. Freeland[4], Dongwon Shin[1], Hiromichi Ohta[6], Matthew F. Chisholm[1] and Ho Nyung Lee[1]\***

[1]Materials Science and Technology Division, Oak Ridge National Laboratory, Oak Ridge, Tennessee 37831, USA
[2]Center for Nanophase Materials Science, Oak Ridge National Laboratory, Oak Ridge, Tennessee 37830, USA
[3]Materials Science Division, Argonne National Laboratory, Argonne, Illinois 60439, USA
[4]Advanced Photon Source, Argonne National Laboratory, Argonne, Illinois 60439, USA
[5]Department of Materials Science and Engineering, Northwestern University, Evanston, Illinois 60208, USA
[6]Research Institute for Electronic Science, Hokkaido University, Sapporo 001-0020, Japan
*E-mail: hnlee@ornl.gov

**Fast, reversible redox reactions in solids at low temperatures without thermomechanical degradation are a promising strategy for enhancing the overall performance and lifetime of many energy materials and devices. However, the robust nature of the cation's oxidation state and the high thermodynamic barrier have hindered the realization of fast catalysis and bulk diffusion at low temperatures. Here, we report a significant lowering of the redox temperature by epitaxial stabilization of strontium cobaltites (SrCoO$_x$) grown directly as one of two distinct crystalline phases, either the perovskite SrCoO$_{3-\delta}$ or the brownmillerite SrCoO$_{2.5}$. Importantly, these two phases can be reversibly switched at a remarkably reduced temperature (200~300 °C) in a considerably short time (< 1 min) without destroying the parent framework. The fast, low temperature redox activity in SrCoO$_{3-\delta}$ is attributed to a small Gibbs free energy difference between two topotatic phases. Our findings thus provide useful information for developing highly sensitive electrochemical sensors and low temperature cathode materials.**

Transition-metal oxides (TMOs) have been at the core of cutting-edge information and energy technologies based on their intriguing physical properties. In particular, owing to the high ionic and electronic conductivity offered from the flexibility of transition metal's charge states, multivalent TMOs have attracted attention for potential applications as catalysts[1-5]. Their operation is generally based on the formation of oxygen off-stoichiometry, which is generally achieved by annealing in reducing conditions. Interestingly, an oxygen vacancy ordered structure that can be topotactically transformed, i.e. oxygen composition can be varied without losing the crystallographic orientation and lattice structure of the parent phase, may be particularly beneficial for lowering the redox temperature and useful for understanding the oxygen diffusion mechanism and anisotropic ionic conduction behaviour in layered materials[6-8]. However, only few detailed studies have been conducted on such materials systems. As a result, understanding the phase-conversion processes and the usefulness of well-ordered oxygen vacancy channels for ionic transport remain unknown.

In this context, strontium cobaltite, SrCoO$_x$, is an excellent platform for the study of reversible redox activity and associated property changes, as it exhibits highly contrasting electronic and magnetic ground states depending on the Co oxidation state, reflecting the crucial role of the oxygen content in determining the physical properties of TMOs[9,10]. However, the possibility of reversible redox reaction and the physical properties of SrCoO$_{3-\delta}$ have not been systematically studied due to the difficulty in synthesizing single crystalline materials. The challenge is driven by the high energy of the Co$^{4+}$ valence state, which prefers to convert to the more stable Co$^{2+}$ and/or Co$^{3+}$ states. Therefore, while the SrCoO$_{2.5}$ phase with ordered oxygen vacancy channels is rather readily grown, preparation of the perovskite $SrCoO_{3-\delta}$ phase has been achieved only through two-step processing of the brownmillerite $SrCoO_{2.5}$ phase[9,11-13], which may severely deteriorate the quality of sample, yielding micro-cracks, grain boundaries, and/or impurity phases. Even though reversible topotatic phase changes from $SrCoO_{2.5}$ to $SrCoO_{3-\delta}$ and vice versa were reported with polycrystalline bulk samples[12,13], it was accomplished only in a liquid electrolyte-based electrochemical cell. The gas-phase oxidation process required a long annealing time ($t \geq 10$ hours), despite the high oxygen pressure used for the phase transformation of bulk materials[14]. It is also worthy to note that most perovskites are thermodynamically robust, so that the formation of oxygen off-stoichiometric phases requires a very high temperature (> 700 °C)[15-17]. Thus, the slow chemical conversion and high temperatures required for reversible redox processes are impractical for many technological applications[6].

In this paper, we report epitaxial stabilization of single crystalline brownmillerite and perovskite films. Not only do they exhibit drastically different structural, electronic, and magnetic properties, but also show a relatively fast, reversible phase transformation at a fairly low temperature without deterioration of the crystal structure.

Epitaxial $SrCoO_{2.5}$ and $SrCoO_{3-\delta}$ thin films were grown using pulsed laser epitaxy (PLE). Various substrates, including (001) $SrTiO_3$ (STO) and (001) $(LaAlO_3)_{0.3}$-$(SrAl_{0.5}Ta_{0.5}O_3)_{0.7}$ (LSAT), were used to find the optimal growth condition. While we could readily stabilize $SrCoO_{2.5}$ in a simple molecular oxygen atmosphere, the growth of $SrCoO_{3-\delta}$ epitaxial films required an extremely oxidizing environment similar to those found for Fe-based perovskite oxides[18]. Thus, we used a mixed gas of ozone and oxygen for the direct epitaxy of conducting perovskites. Figures 1a and b show x-ray diffraction (XRD) $\theta$-$2\theta$ scan patterns of a typical brownmillerite-type $SrCoO_{2.5}$ thin film grown on a (001) STO substrate and a perovskite-type $SrCoO_{3-\delta}$ thin film grown on a (001) LSAT substrate, respectively. While both phases could be epitaxially grown on both STO ($a_c$ = 3.905 Å) and LSAT ($a_c$ = 3.868 Å) substrates, we found that higher quality $SrCoO_{2.5}$ and $SrCoO_{3-\delta}$ films could be grown respectively on STO and LSAT due to the low lattice mismatches. [Note that $SrCoO_{2.5}$ is orthorhombic ($a_o$ = 5.5739, $b_o$ = 5.4697, and $c_o$ = 15.7450 Å)[19], which can be represented as pseudo-tetragonal ($a_t$ = 3.905 and $c_t/4$ = 3.9363 Å). $SrCoO_{3-\delta}$ is cubic with $a_c$ = 3.8289 Å[9].] We use the orthorhombic notation throughout the paper. The XRD $\theta$-$2\theta$ scans from both phases reveal well-defined peaks and distinct thickness fringes, which indicate chemically sharp interfaces and flat surfaces. X-ray rocking curve scans further bear out the excellent crystallinity ($\Delta\omega$ < 0.05°), and reciprocal space mapping confirms that both films are coherently strained on the substrates (See Fig. S1). In particular, the $SrCoO_{2.5}$ film clearly exhibits the characteristic doubling of the $c$-axis lattice constant that cannot be found in $SrCoO_{3-\delta}$, originating from the alternate stacking of octahedral and tetrahedral sub-layers along the $c$-axis (film normal) (see the inset in Fig. 1a for a sketch of the brownmillerite crystal structure). The direct synthesis of the perovskite phase, without post-growth oxidation, has not been reported so far, thus, this demonstration of epitaxial $SrCoO_{3-\delta}$ thin film growth should stimulate further studies on oxide synthesis.

Figure 1c shows cross-sectional Z-contrast scanning transmission electron microscopy (Z-STEM) images of a $SrCoO_{2.5}$ film seen along the [110] STO direction. Here, the alternate stacking of fully oxygenated octahedral and oxygen-deficient tetrahedral sub-layers is observed directly, and is consistent with the brownmillerite structure. No impurity phases or grain boundaries, which usually include cobalt precipitates[20], were observed, confirming the XRD results. Most importantly, local structural changes due to the oxygen deficiency were clearly visualized: (1) Co–Co lateral atomic spacing modulation within the tetrahedral layers owing to the volume expansion near oxygen vacancy sites; (2) ~30% increase in the sub unit-cell height (e.g. vertical Sr–Sr atomic spacing) of the oxygen deficient tetrahedral layers ($c'/2$ = 4.3 ± 0.1 Å) over that of the octahedral layers ($c''/2$ = 3.4 ± 0.1 Å); and (3) local spatial modulation of lateral Sr–Sr atomic spacing (2.8 ± 0.1 and 2.6 ± 0.1 Å), which is induced by a local tilt of octahedra. We note that the octahedral tilt in $SrCoO_{2.5}$ has never been visualized previously at the atomic scale. Such collective displacements of Co ions in the tetrahedral layers create well-ordered 1D vacancy channels, which form a zig-zag network on the (010) plane of $SrCoO_{2.5}$, as shown in the inset of Fig. 1a. Even though it is not the scope of the current work, such well-ordered vacancy channels offer an interesting playground for the study of fast ion conductivity because it is known that good ionic conductors contain large open frameworks[4,21].

In order to unambiguously confirm the two chemically distinct phases, we investigated the element-resolved details of the Co valence state of epitaxial $SrCoO_{2.5}$ and $SrCoO_{3-\delta}$ thin films on STO and LSAT, respectively, by polarized x-ray absorption spectroscopy (XAS). This technique provides information on the oxidation state of Co, which plays a deterministic role in the magnetic, electronic, and catalytic properties of the materials[22-24]. Oxygen stoichiometry in our $SrCoO_{2.5}$ and $SrCoO_{3-\delta}$ thin films was characterized by XAS at the Co $L$-edge, as well as by the pre-peak at the O $K$-edge, which is related to the Co $3d$ –O $2p$ hybridization[25-27], as shown in Fig. 2a. It has been reported that the pre-peak (~527 eV) intensity of the O $K$-edge decreases substantially as $\delta$ approaches 0.5 in polycrystalline $SrCoO_{3-\delta}$[27]. The weak pre-peak observed for the $SrCoO_{2.5}$ films is consistent with a $Co^{3+}$ valence state, while the relative intensity of the pre-peak for the $SrCoO_{3-\delta}$ film is greater than that for polycrystalline $SrCoO_{2.82}$. The oxygen stoichiometry was more rigorously determined from the XAS measurement of the Co $L$-edge (Fig. 2b), which was compared with a bulk reference of known oxygen stoichiometry ($SrCoO_{2.88}$), as determined by thermogravimetric analysis[28], as shown in Fig. S2. Together with the clear metallic behaviour that is a good indicator of the oxygen stoichiometry to be beyond $SrCoO_{2.90}$ (Ref. [29]) as shown in Fig. 3c and based on the shift of Co $L_2$-edge peak from bulk $SrCoO_{2.88}$, we could determine the highly oxygenated state of $SrCoO_{3-\delta}$ to be $\delta \leq 0.1$ (Ref. [27]). In prior studies, achieving such a highly oxygenated crystalline strontium cobaltite has required post-treatment[9,11,29,30].

The distinct chemical valence difference between the $SrCoO_{2.5}$ and $SrCoO_{3-\delta}$ phases produced distinct magnetic properties. Element-resolved measurements of the net magnetic moment using x-ray magnetic circular dichroism (XMCD) indeed showed a large ferromagnetic signal in the $SrCoO_{3-\delta}$ film (see Fig. 2c). The ferromagnetic state was further supported by the field-dependent XMCD data revealing that about 70% of the XMCD signal at 5 T was retained at 0.1 T. On the other hand, the $SrCoO_{2.5}$ displayed hardly any XMCD signal even at 5 T, consistent with an antiferromagnetic ground state. Furthermore, as shown in Figs. 3a and b, the magnetization measurements confirmed the XMCD results. Figure 3a shows that our $SrCoO_{3-\delta}$ epitaxial films on LSAT are indeed ferromagnetic below ~250 K. The Curie temperature ($T_C$) is slightly lower than that of a $SrCoO_{3-\delta}$ single crystal (~305 K)[9]. The lower $T_C$ is due most probably to the substrate induced tensile strain[31]. As shown in Fig. 3b, the saturation magnetization ($M_S$) of $SrCoO_{3-\delta}$ at 10 K was ~2.3 $\mu_B$/Co, slightly smaller than the bulk value with an intermediate spin state that can be attributed to tensile strain as well. The magnetization data experimentally confirm the theoretically predicted spin state in $SrCoO_{3-\delta}$ (Ref. 32). In contrast to $SrCoO_{3-\delta}$, no significant SQUID signal was observed from the $SrCoO_{2.5}$ film as shown in Figs. 3a and b. Moreover, the $M(H)$ curves of $SrCoO_{2.5}$ recorded at 10 and 250 K (data not shown) did not show any discernible differences, supporting the fact that our $SrCoO_{2.5}$ epitaxial films are indeed antiferromagnetic.

In addition to the magnetic ground states in these epitaxial thin films, we observed a considerable difference in the electronic transport properties of both films. There is a large change in resistivity between the two phases – we observed a more than four orders of magnitude difference in resistivity between $SrCoO_{2.5}$ and $SrCoO_{3-\delta}$ even at room temperature (Fig. 3c). A clear observation of metallic ground state in our $SrCoO_{3-\delta}$ film indeed represents the successful stabilization of $Co^{4+}$ in our films, as the insulator-to-metal transition occurs at $\delta \sim 0.1$[9,29]. It is worth noting that our $SrCoO_{3-\delta}$ showed a highly lower resistivity than that of polycrystalline samples[11,29], indicating minimal oxygen deficiency and excellent film quality from the direct epitaxy. In contrast, $SrCoO_{2.5}$ films were highly insulating (see Fig.

3c). We further found that, regardless of the type of substrates, highly insulating brownmillerite $SrCoO_{2.5}$ and highly metallic perovskite $SrCoO_{3-\delta}$ films could be grown *in situ* under well-optimized growth conditions. In addition, since the thermopower (*S*) reflects the electronic ground state, i.e. insulator or metal, we have compared the *S*-value of $SrCoO_{2.5}$ with that of $SrCoO_{3-\delta}$ at room temperature. As shown in Fig. 3d, the *S*-value of $SrCoO_{2.5}$ ($S = +254$ $\mu V/K$) is significantly greater than that of $SrCoO_{3-\delta}$ ($S = +9.6$ $\mu V/K$), confirming the insulating nature of $SrCoO_{2.5}$. It is worth mentioning that the observed *S*-values of both $SrCoO_{3-\delta}$ and $SrCoO_{2.5}$ are positive, which indicates the *p*-type conduction, consistent with bulk results[29,32-34]. These results clearly indicate the high sensitivity of electronic transport properties to changes in the oxygen content.

As mentioned earlier, the realization of oxygen content control is of significant interest for both the underlying physics and technological applications of multivalent oxides. Therefore, we examined the feasibility of reversible redox activities in our epitaxial strontium cobaltites without destroying the parent structure. In order to directly observe the redox reactions, we monitored the phase reversal processes by *real-time* recording of temperature-dependent XRD $\theta$-$2\theta$ scans with epitaxial films on LSAT in vacuum and at oxygen atmosphere. As shown in Fig. 4a for the reduction process (i.e., $SrCoO_{3-\delta}$-to-$SrCoO_{2.5}$ transformation), the 002 peak from the $SrCoO_{3-\delta}$ film starts to disappear at 175 °C, and then a complete transition to $SrCoO_{2.5}$, which can be confirmed by the elongated *c*–axis lattice constant due to the oxygen vacancy ordering, is observed at 210 °C. The resulting *c*-axis orientation, i.e. oxygen vacancy channels aligned parallel to the interface in the oxygen reduced film on LSAT, can be understood by the lower mismatch (lattice mismatch = 0.96%) as compared to the (110)-orientation (lattice mismatch = 1.76% along the *c*-axis) with the vacancy channels running out of plane. For the oxidation process (i.e., $SrCoO_{2.5}$-to-$SrCoO_{3-\delta}$ transformation), a complete oxidation could be achieved at around 350 °C in 5 bar of $O_2$, as shown in Fig. 4b. As shown in Figs. 3c and S3, the samples obtained through the rapid topotactic phase conversion revealed overall similar physical properties as compared to those from the direct growth. However, the topotatic phase conversion of a brownmillerite film to a highly metallic perovskite film was a complicated process, and the specifics of which did depend on the type of substrates (e.g., the condition for the phase conversion of thin films on LSAT was less stringent than that on STO due to a lower lattice mismatch with the perovskite phase) and details of the annealing condition (e.g., pressure, time, and temperature). Nonetheless, we stress that the topotactic phase conversion at low temperature remains a highly effective way to stabilize highly oxidized perovskite films as shown in Fig. 3c, and the high crystallinity of our single crystalline films is highly beneficial to improve the completeness of the phase conversion.

In addition to the surprisingly low temperature for the topotactic phase transformation process (i.e., oxygen content change), the phase conversion was notably fast (the entire XRD scanning for the phase conversion took less than 10 minutes as described in Methods), as compared to previous reports (from ~10 h or to one week) by conventional high pressure annealing at similar temperatures or by room temperature electrochemical approaches[11,13,14,30]. Moreover, the oxidation pressure is at least several hundred times lower than the annealing approach previously reported[14]. We further note that the conversion time and oxygen pressure for our epitaxial thin films could be reduced to one minute and 0.67 bar, respectively. The lower limits of the pressure and time were determined by a step-by-step change of the post-annealing temperature right after the growth without breaking the vacuum (data not shown).

It is interesting to compare the topotactic phase conversion of $SrCoO_x$ thin films with the ferrites as the latter materials are also known to show similar phase evolution. Pioneering studies have been conducted with the $CaFeO_x$ and $SrFeO_x$ systems,[8,35] where topotactic phase changes from perovskite to not only brownmillerite, but also the thermodynamically less stable, infinite layered square-planar structure ($x = 2.0$) were observed at low temperatures when treated with strong oxygen reducers. These observations also provided crucial information regarding the possibility of topotactic valence changes with active control of the mobile lattice oxygen with chemical reducers. However, the possibility of their phase reversal to the perovskite form by oxidation has not been much explored. Thus, we focused on understating the thermodynamic phase stability of $SrCoO_x$ at low temperatures by comparing with other multivalent oxides. Since the phase stability of the bulk $SrCoO_{2.5}$-$SrCoO_3$ pseudo-binary systems have not been comprehensively studied, we used a computational thermodynamic approach to quantify the energy barrier for the phase transition. The calculated Gibbs free energy difference between the brownmillerite $SrCoO_{2.5}$ and the perovskite $SrCoO_{3-\delta}$ at different oxygen contents ($0 < \delta < 0.3$) as a function of temperature is represented in Fig. 5. We compared the energy barrier with that of the widely studied manganites, i.e. $SrMnO_x$, which also forms $SrMnO_{2.5}$ and $SrMnO_{3-\delta}$ phases. Interestingly, the overall magnitude of the energy difference between $SrCoO_{2.5}$ and $SrCoO_{3-\delta}$ is smaller, in particular at low temperatures, than that of $SrMnO_x$ at a given oxygen content. It is noteworthy that the energy difference between $SrCoO_{2.5}$ and $SrCoO_{3-\delta}$ decreases significantly (at least 30%) with a small deviation (e.g., $\delta = 0.1$) from perfect stoichiometry, while the maximum change in $SrMnO_x$ is only 20% at the same oxygen content. This significantly reduced energy barrier is responsible for the low temperature phase transition between $SrCoO_{2.5}$ and $SrCoO_{3-\delta}$ at low temperatures. This is not the case for $SrMnO_x$, in which the energy difference between the $SrMnO_{2.5}$ and $SrMnO_{3-\delta}$ phases remains large regardless of the change in oxygen stoichiometry and temperature. These thermodynamic considerations provide considerable advantages of $SrCoO_x$ over other conventional perovskites for rapid topotatic phase control. Moreover, as one can observe in Fig. 5, the formation of $SrCoO_{2.5}$ is more energetically favourable than $SrCoO_{3-\delta}$, whereas $SrMnO_{3-\delta}$ is more favourable than $SrMnO_{2.5}$. Interestingly, the Gibbs free energy difference for highly oxygenated $SrCoO_{3-\delta}$ ($\delta < 0.3$) increases as the temperature increases, while it is opposite for $SrMnO_x$. Therefore, our finding on the thermodynamic energy difference in $SrCoO_x$ provides crucial information needed for understanding the topotatic processes in multivalent cobaltites.

Finally, we note that the fast, reversible redox activity implies high catalytic activity at relatively low temperatures. To elucidate the potential of $SrCoO_x$ as a heterogeneous catalyst, we studied its activity for carbon monoxide oxidation, a typical probe reaction. Due to the extremely small surface areas of the epitaxial films, a custom designed micro-reactor was employed using inlet gas streams of CO (0.1 mbar) and $O_2$ (0.1 mbar). The effluent gas mixture was analyzed in-line with a gas-chromatograph and mass spectroscopy detector (see Methods for details). For this reaction, the oxygen activity in the reactor is low; thus, the epitaxial $SrCoO_{3-\delta}$ film was unstable and the epitaxial $SrCoO_{2.5}$ film on LSAT was chosen for the catalytic study. Here the presence of catalytic activity can be connected to both the conversion of CO and the production of $CO_2$. The conversion of the inlet CO gas is shown in Fig. 6a. As compared with the clean reactor (no sample), clear conversion of CO was observed above ~320 °C, increasing at a significant rate. At the same temperatures, we observed an uptake in $CO_2$ production as shown in Fig. 6b. To our knowledge this is the first evidence for catalytic activity from the $SrCoO_{2.5}$ phase. Furthermore, since substantial activity can be measured at relatively low temperatures from a sample with a surface area of only 0.5 $cm^2$, this suggests that stabilized $SrCoO_{2.5}$ has good potential as a catalyst for many other important redox reactions.

In conclusion, we found that rapid, reversible redox activity could be accomplished at significantly reduced temperatures (as low as 200 °C) from epitaxially stabilized, oxygen vacancy ordered $SrCoO_{2.5}$ and thermodynamically unfavourable perovskite $SrCoO_{3-\delta}$ single crystalline thin films. In addition, these topotactic phases have the sharply distinctive magnetic and electronic properties (i.e. antiferromagnetic *vs*. ferromagnetic and insulating *vs*. metallic). More interestingly, the $SrCoO_{2.5}$ film exhibited significant activity for the carbon monoxide oxidation reaction, which suggests good potential for $SrCoO_{2.5}$ as a redox catalyst. The excellent materials performance we reported here, therefore, provides a pathway to the design of new ionic materials, in which the dynamic change in the oxidation state plays a pivotal role for energy generation, storage, and electrochemical sensing.

**Methods**

Epitaxial $SrCoO_{2.5}$ and $SrCoO_{3-\delta}$ thin films (30-60 nm in thickness) were grown on (001) STO and (001) LSAT substrates by PLE (KrF, λ = 248 nm). All the films were grown at 750 °C in 0.013 mbar of $O_2$ for the $SrCoO_{2.5}$ and 0.267 mbar of $O_2 + O_3$ (5%) for the $SrCoO_{3-\delta}$. The laser fluence was fixed at 1.7 J/cm². The sample structure and crystallinity were characterized by high-resolution four-circle XRD (X'Pert, Panalytical Inc.). The Z-contrast images were obtained using a Nion UltraSTEM 200 operated at 200 keV.

The valence (or oxidation) state and magnetism in $SrCoO_x$ were elucidated by XAS and XMCD at beamline 4-ID-C of the Advanced Photon Source, Argonne National Laboratory. Magnetic properties were characterized with a 7 T Superconducting Quantum Interference Device (SQUID) magnetometer (Quantum Design). Temperature dependent DC transport measurements with van der Pauw geometry were performed with a 14 T Physical Property Measurement System (PPMS) (Quantum Design). The thermopower values were also measured by a conventional steady state method using two Peltier devices under the thin films to give a temperature difference ($\Delta V$~10 K). To eliminate the different substrate contributions, both $SrCoO_{3-\delta}$ and $SrCoO_{2.5}$ phases were grown on (001) STO substrates.

The reversible redox reactions were monitored by using high temperature environmental XRD, in which gas type, flow rate, and pressure can be controlled. For the $SrCoO_{3-\delta}$-to-$SrCoO_{2.5}$ conversion, a high-resolution four-circle XRD (X'Pert, Panalytical Inc.) with a domed hot stage (DHS 900, Anton Paar) was used. The inside of dome was evacuated with a mechanical pump to a base pressure of 0.0013 mbar. For the $SrCoO_{2.5}$-to-$SrCoO_{3-\delta}$ conversion, a powder XRD (Panalytical Inc.) with a reactor chamber (XRK 900, Anton Paar) was used to pressurize the inside of the heating chamber (5 bar of $O_2$). The ramping rate was 30-60 °C/min and the average scan time was 2-2.5 minutes.

Thermodynamic descriptions for $SrCoO_{2.5}$ and $SrCoO_{3-\delta}$ were taken from the thermodynamic modeling of $SrCoO_x$, whose model parameters were critically and self-consistently evaluated to reproduce both phase equilibrium and thermochemistry data for their bulk form. All the equilibrium phases in the thermodynamic modeling of $SrCoO_x$ other than $SrCoO_{2.5}$ and $SrCoO_{3-\delta}$ were suspended in the present thermodynamic calculation to compute energy difference only between $SrCoO_{2.5}$ and $SrCoO_{3-\delta}$. Thermodynamic descriptions for the $SrCoO_{2.5}$ and $SrCoO_{3-\delta}$ were obtained from the literature[36]. The $SrCoO_{2.5}$ and $SrMnO_{2.5}$ were modeled as stoichiometric, while the perovskite phases were modeled as solution phases. The designated oxygen sublattices for perovskites allow mixing between oxygen and vacancy to achieve hypo-stoichiometry.

Gas-phase catalysis measurements were made with a custom micro-reactor with a volume < 50 ml. The high level of reactor cleanliness was achieved by limiting material within the reactor to fused quartz, stainless steel and fluoropolymer seals. Moreover, silver paint used for mounting the substrate for the film growth was completely removed before inserting the $SrCoO_{2.5}$ thin film into the reactor for the catalysis experiment. Heating was conducted by passing light from a halogen bulb through a fused quartz platform to the backside of the sample. The inlet gas streams consisted of 300 ppm CO and 300 ppm $O_2$, both mixed with a He balance (resulting in a partial pressure of 0.1 mbar for each gas), and a throughput of 5 SCCM. The temperature was programmed at 30 °C steps, each held for 16 min. At each step, the initial 10 min were reserved to allow the system to reach steady state conditions. Afterwards, an aliquot of the gas stream (20 μl) was injected into the gas-chromatograph / mass spec (Perkin Elmer®). A carbon packed capillary column was used to separate the CO and residual $N_2$ in the gas sample. The total ion chromatogram was mass separated to isolated peaks from CO and $CO_2$ and then integrated. The concentration was determined with the integrated values and calibrated values from known gas mixtures. Detection limits for the CO conversion were set by the surfaces within the reactor, as determined by running the temperature program without a sample loaded and measuring the CO levels. In contrast, the detection limit for $CO_2$ was set by the sensitivity of the mass spectrometer (~ 4 ppb). The reproducibility of the conversion trends was substantiated with measurement of an additional $SrCoO_{2.5}$ epitaxial thin film.

## Acknowledgements
The work was supported by the U.S. Department of Energy, Basic Energy Sciences, Materials Sciences and Engineering Division. The *in situ* XRD measurement was conducted at the Center for Nanophase Materials Sciences, which is sponsored at Oak Ridge National Laboratory by the Scientific User Facilities Division, Office of Basic Energy Sciences, U.S. Department of Energy. Use of the Advanced Photon Source was supported by the U. S. Department of Energy, Office of Science, under Contract No.DE-AC02-06CH11357. HO was supported by MEXT (25246023).

## Author contributions
H.J. conducted sample synthesis, XRD, DC transport, SQUID measurements with help of W.S.C, and H.J. and M.C.B. performed the high temperature environmental XRD, under the direction of H.N.L. M.F.C. performed STEM measurements. I.C.T. and J.W.F. measured XAS and XMCD, and H.O. worked on thermopower measurements. D.S. performed the thermodynamic modelling. C.M.F. and D.D.F. worked on the catalysis measurement. H.N.L. initiated the research and supervised the work. All authors participated in writing the manuscript.

## Competing Financial Interests statement
The authors declare no competing financial interests.

## Additional information
Supplementary information accompanies this paper on www.nature.com/naturematerials. Reprints and permissions information is available online at http://www.nature.com/reprints. Correspondence and requests for materials should be addressed to H.N.L.

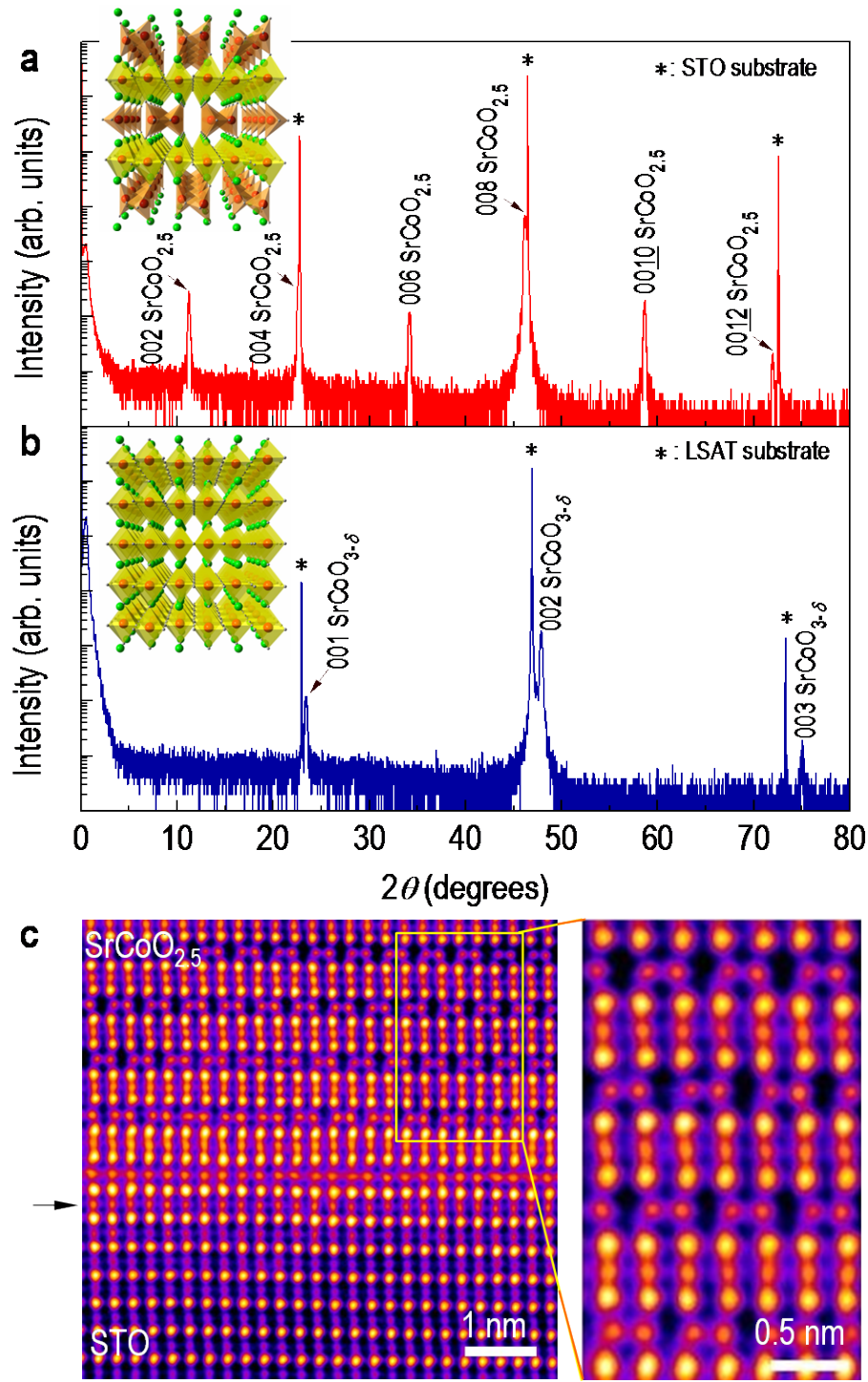

**Figure 1 | Epitaxial synthesis of two topotatic SrCoO$_x$ phases. a,b,** X-ray diffraction $\theta$-$2\theta$ scans of **(a)** a brownmillerite SrCoO$_{2.5}$ film on (001) STO and **(b)** a perovskite SrCoO$_{3-\delta}$ film on (001) LSAT. Insets are schematics of SrCoO$_{2.5}$ and SrCoO$_3$. **c,** Cross-sectional Z-contrast STEM image of a SrCoO$_{2.5}$ film on STO along the [110] STO direction. The arrow indicates the interface between film and substrate. A magnified image of SrCoO$_{2.5}$ on the right-hand side is shown to clearly visualize the 1D oxygen vacancy channels and subsequent local structural distortions.

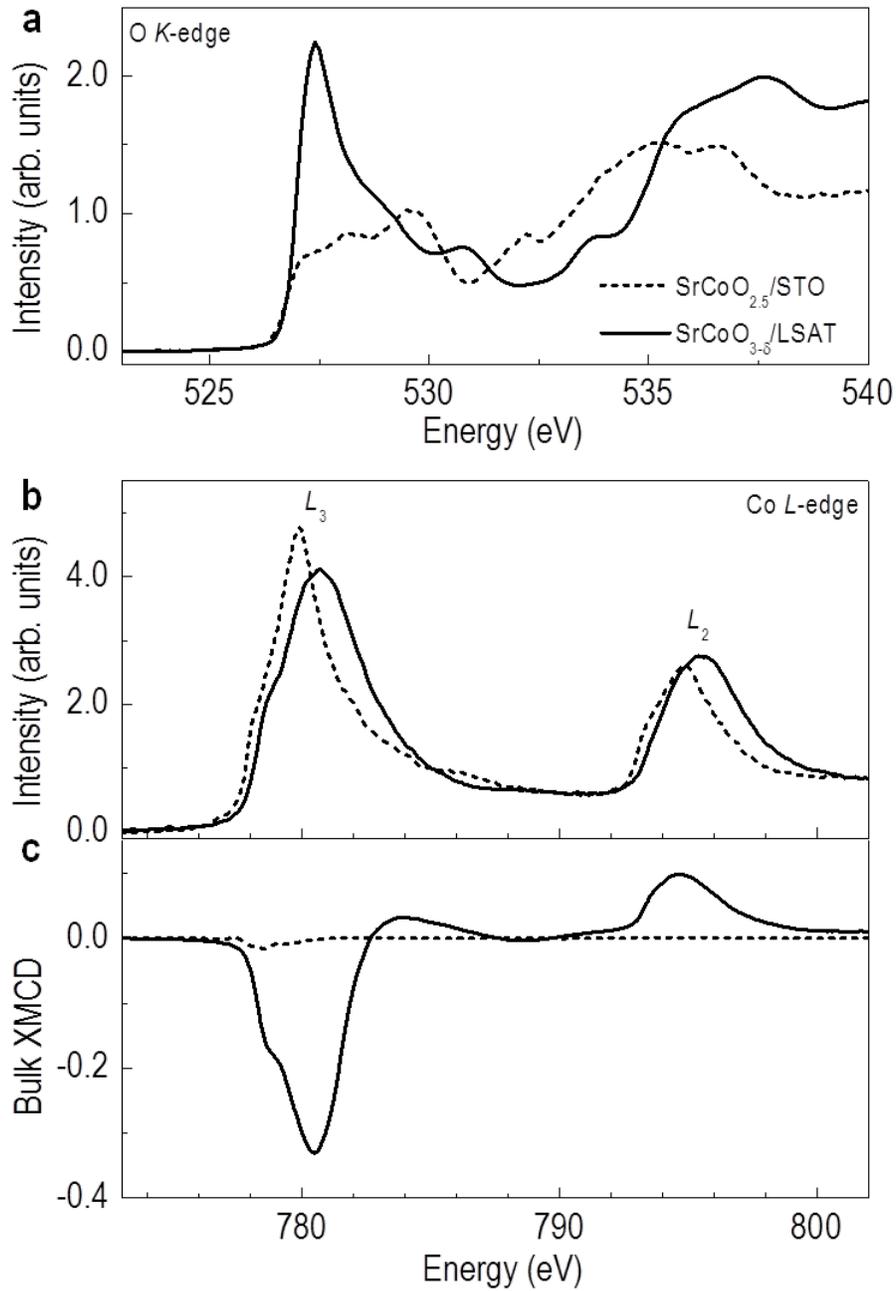

**Figure 2 | Comparison of oxidation states and magnetism. a,** XAS O *K*-edge spectra of SrCoO$_{3-\delta}$ and SrCoO$_{2.5}$ films on LSAT and STO, respectively. A clear pre-peak at around 527 eV clearly indicates different oxygen contents in SrCoO$_{3-\delta}$ (solid line) and SrCoO$_{2.5}$ (dashed line) films. **b,** XAS Co $L_{2,3}$-edge spectra. The shift of $L_3$-edge towards the higher energy (> 0.7 eV) in SrCoO$_{3-\delta}$ indicates the Co ions in SrCoO$_{3-\delta}$ are in a higher valence state. **c,** XMCD spectra of the two phases at 5 T.

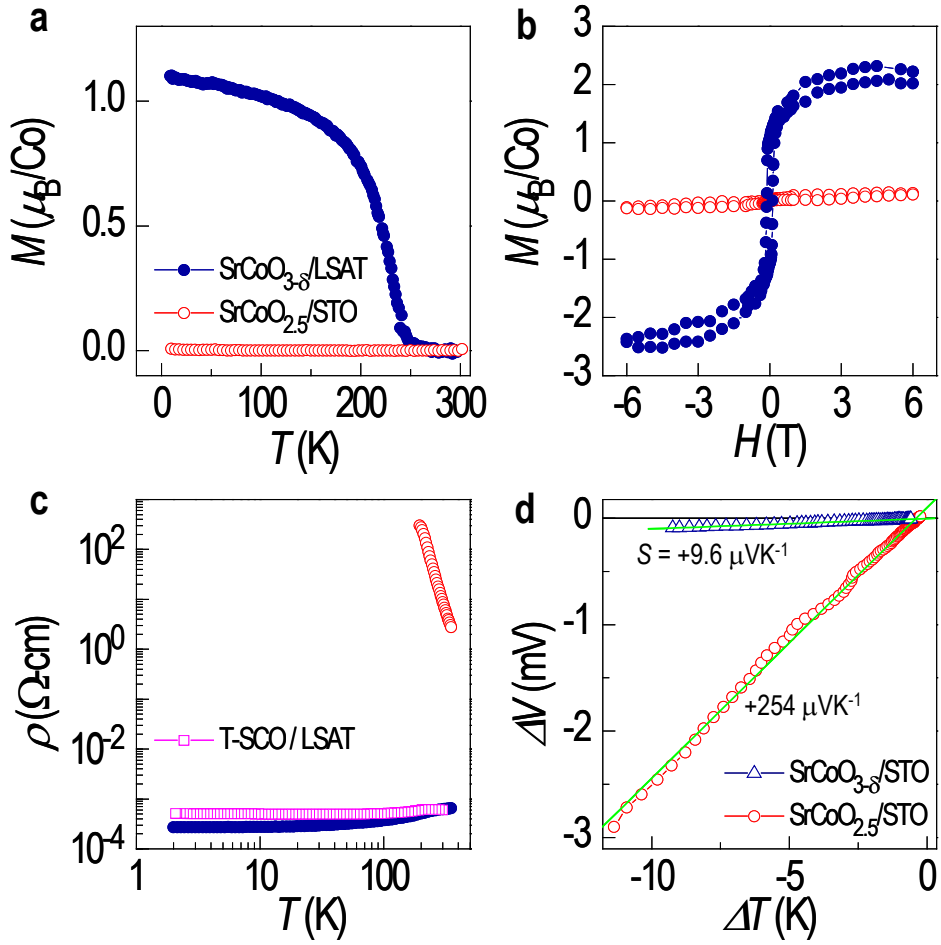

**Figure 3 | Magnetic and *dc* transport properties. a,b,** (a) Temperature dependent magnetization of $SrCoO_{2.5}$ and $SrCoO_{3-\delta}$ thin films at 1000 Oe and **(b)** magnetic hysteresis loops at 10 K. **c.** Resistivity of $SrCoO_{2.5}$ and $SrCoO_{3-\delta}$ thin films as a function of temperature, clearly showing insulating and metallic states, respectively. The $\rho(T)$ curve marked with T-SCO is obtained from a perovskite film, topotactically-oxidized from a brownmillerite film by annealing at 300 °C in 0.67 bar of $O_2$ for 5 min., confirming the successful phase conversion based on the clear metallic behaviour similarly seen from the *in situ* grown perovskite film. **d.** Thermoelectromotive force ($\Delta V$) at 300 K with a strong contrast in the thermopower (*S*) between the two phases.

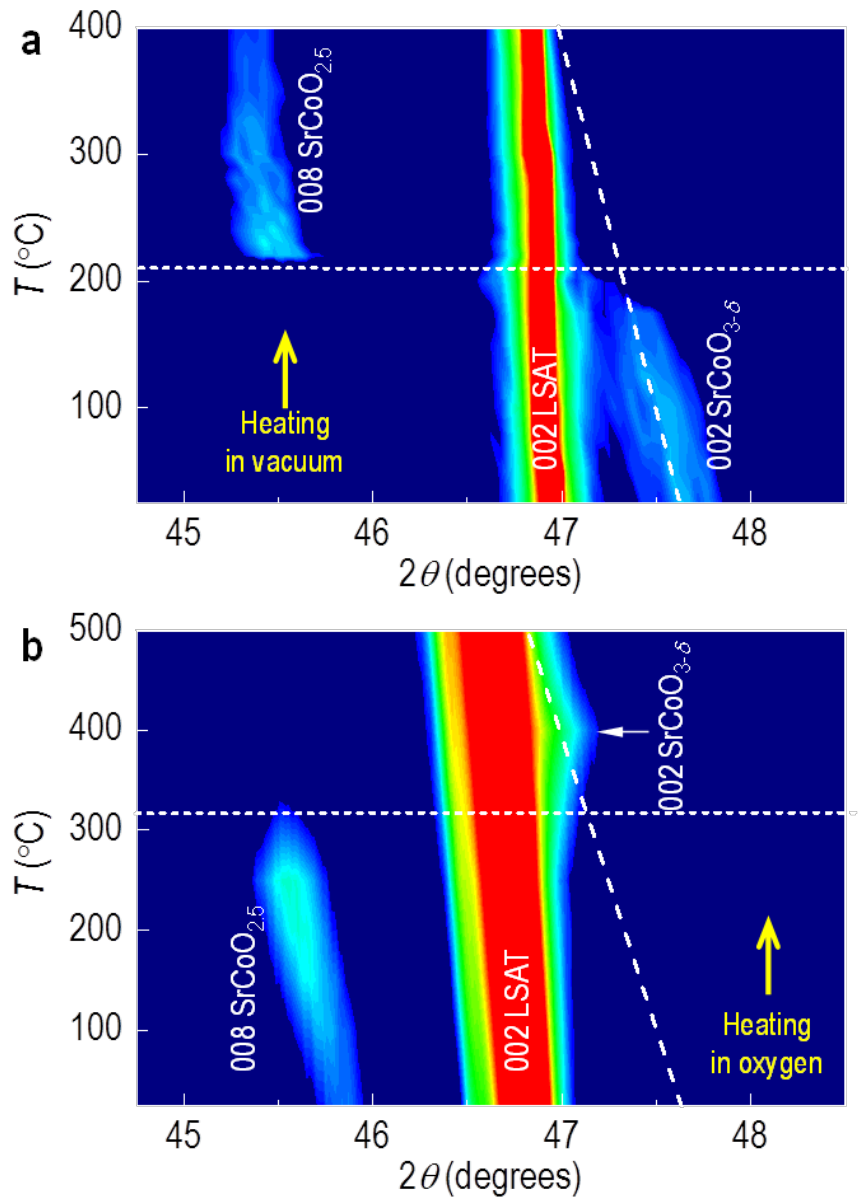

**Figure 4 | Direct probing of reversible redox activity. a, b,** Real-time temperature dependent XRD $\theta$-$2\theta$ scans around the 002 LSAT reflection, clearly revealing **(a)** the $SrCoO_{3-\delta}$-to-$SrCoO_{2.5}$ transition (reduction) in vacuum and **(b)** the $SrCoO_{2.5}$-to-$SrCoO_{3-\delta}$ transition (oxidation) in oxygen. The $SrCoO_{3-\delta}$-to-$SrCoO_{2.5}$- transition is at ~210 °C, while the reversal is at ~350 °C. Note that, due to the large difference in the thermal expansion coefficient between the film and substrate, the 002 $SrCoO_{3-\delta}$ film peak overlaps with the substrate one at high temperatures as marked with the dashed lines.

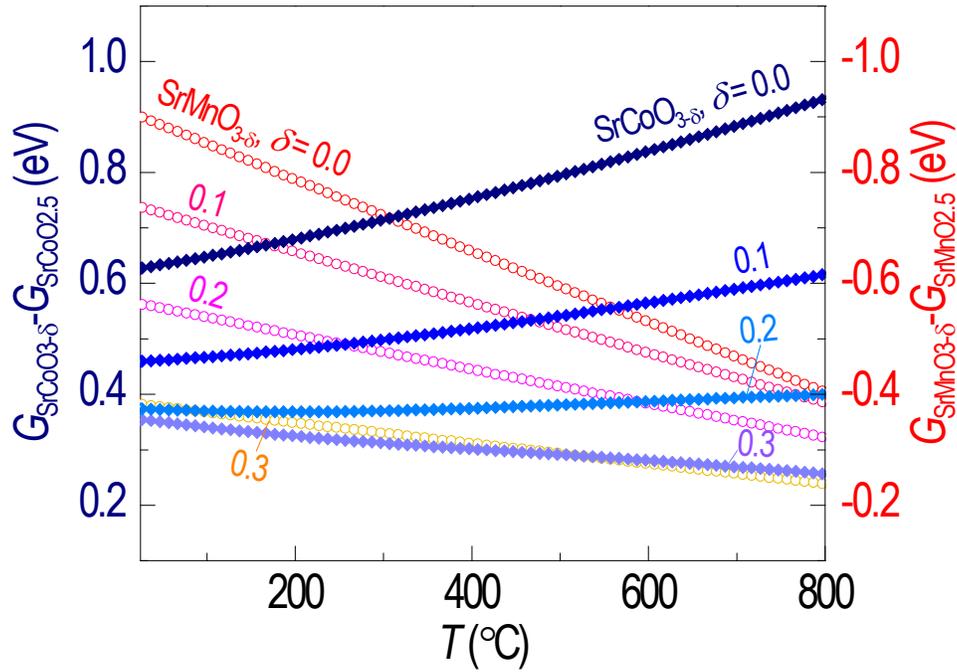

**Figure 5 | Thermodynamic competition.** Gibbs energy differences between SrCoO$_x$ and SrMnO$_x$ with different oxygen contents shown as a function of temperature. Note that SrCoO$_{2.5}$ and SrMnO$_{3-\delta}$ phases are energetically more favourable than the other phases. When the temperature decreases, the energy difference is reduced significantly in case of SrCoO$_x$, facilitating the topotatic phase conversion. The opposite trend is observed in SrMnO$_x$.

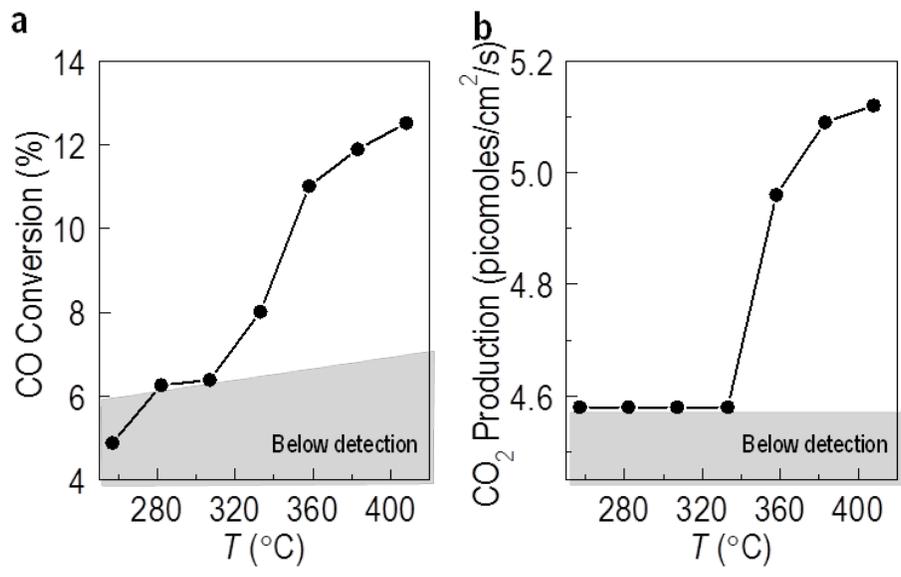

**Figure 6 | Gas phase catalysis. a, b,** Temperature programmed CO oxidation reaction over a brownmillerite film on a LSAT substrate. The CO conversion **(a)** and the $CO_2$ production **(b)** both show catalytic activity above ~320°C.

# Reversible redox reactions in an epitaxially stabilized SrCoO$_x$ oxygen sponge

Hyoungjeen Jeen[1], Woo Seok Choi[1], Michael D. Biegalski[2], Chad M. Folkman[3], I-Cheng Tung[4,5], Dillon D. Fong[3], John W. Freeland[4], Dongwon Shin[1], Hiromichi Ohta[6], Matthew F. Chisholm[1] and Ho Nyung Lee[1,*]

[1]Materials Science and Technology Division, Oak Ridge National Laboratory, Oak Ridge, Tennessee 37831, USA
[2]Center for Nanophase Materials Science, Oak Ridge National Laboratory, Oak Ridge, Tennessee 37830, USA
[3]Materials Science Division, Argonne National Laboratory, Argonne, Illinois 60439, USA
[4]Advanced Photon Source, Argonne National Laboratory, Argonne, Illinois 60439, USA
[5]Department of Materials Science and Engineering, Northwestern University, Evanston, Illinois 60208, USA
[6]Research Institute for Electronic Science, Hokkaido University, Sapporo 001-0020, Japan
*E-mail: hnlee@ornl.gov

## CRYSTALLINITY AND STRAIN STATE IN SCO EPITAXIAL THIN FILMS

We performed a detailed x-ray diffraction study. Figures S1a and b show x-ray rocking curves of the 008 Bragg peak of SrCoO$_{2.5}$ and the 002 Bragg peak of SrCoO$_{3-\delta}$. They revealed full width at half maxima (FWHM) of 0.04º and 0.05º, respectively, confirming the excellent crystallinity. (*cf*., typical FWHM in $\omega$ scan of the 002 STO peak is ~0.015º.) Note that, due to a larger mismatch, even though a SrCoO$_{3-\delta}$ film could be also grown on STO, the crystallinity was poorer, and it was more difficult to oxidize than that on LSAT. In order to characterize the in-plane strain state, we also recorded reciprocal space maps (RSMs) around the 103 STO and LSAT peaks, as shown in Figs. S1c and d, respectively. The in-plane lattice constants of both SrCoO$_{2.5}$ and SrCoO$_{3-\delta}$ films were coherently matched to those of the substrates. In addition, the $c$ lattice constants from the RSMs were consistent with the ones measured from the $\theta$-$2\theta$ scans, i.e. $c/2$ = 3.93 Å for SrCoO$_{2.5}$ and $c$ = 3.79 Å for SrCoO$_{3-\delta}$. The $c$-axis lattice constant of SrCoO$_{3-\delta}$ was smaller than the bulk value due to the substrate induced tensile strain.

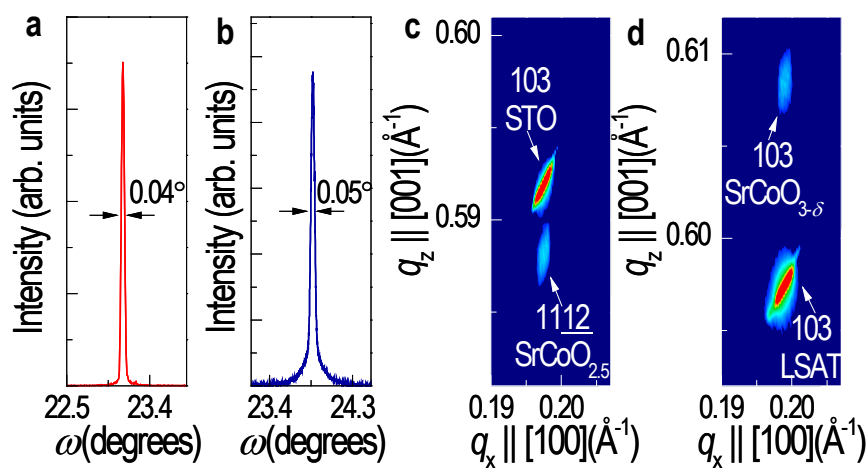

**Figure S1. a,b,** XRD rocking curves of **(a)** the 008 brownmillerite SrCoO$_{2.5}$ peak and **(b)** the 002 of perovskite SrCoO$_{3-\delta}$ peak. **c,d,** Reciprocal space maps of **(c)** a SrCoO$_{2.5}$ thin film on a STO substrate and **(d)** a SrCoO$_{3-\delta}$ thin film on a LSAT substrate.

## OXIDATION STATE BY XAS

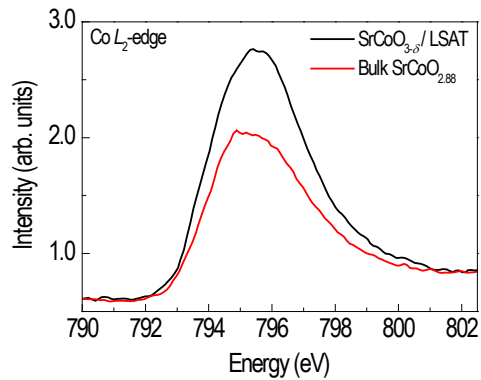

**Figure S2.** XAS Co $L_2$-edge spectra of the SrCoO$_{3-\delta}$ film on (001) LSAT (shown in Fig. 2b) comparatively represented with a bulk sample, confirming the high oxidation state beyond $x = 2.88$.

## RAPID TOPOTACTIC OXIDATION OF SrCoO$_{2.5}$ THIN FIMS AT LOW TEMPERATURES

Figure S3 shows XRD data for topotactic reversal between the brownmillerite and perovskite phases. In order to confirm the oxidation of the brownmillerite phase to the perovskite, we show XRD data recorded at 100 and 30 °C upon cooling. The perovskite 002 peak that overlaps with the substrate peak at high temperatures is marked (note the low peak intensity originates from the small film thickness), showing a clear peak separation.

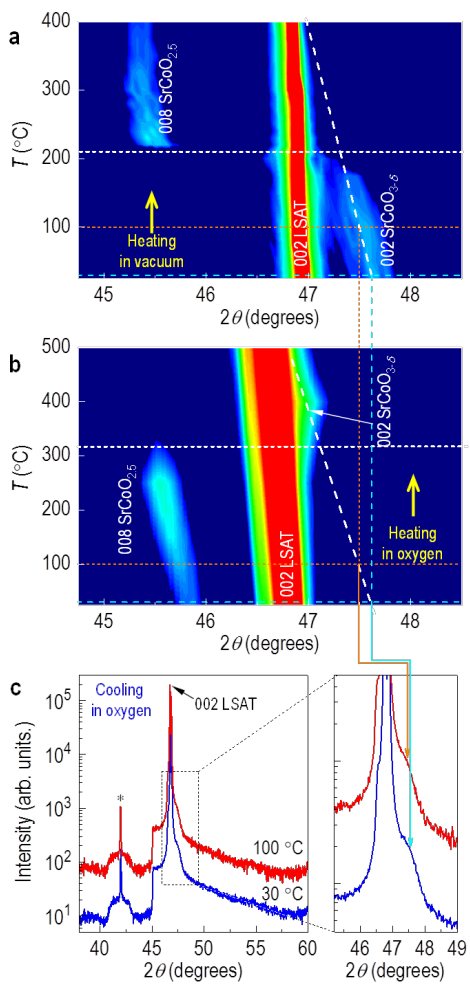

**Figure S3. a,b,** Real-time temperature dependent XRD $\theta$-$2\theta$ scans for (**a**) reduction of SrCoO$_{3-\delta}$ and (**b**) oxidation of SrCoO$_{2.5}$, represented in Fig. 3. The large thermal expansion of SrCoO$_{3-\delta}$ is marked with a white dashed line, which corresponds to $5.6 \times 10^{-5}$ K$^{-1}$ along the out-of-plane direction. Horizontal dotted lines in cyan and orange indicate peak positions at 30 and 100 °C, respectively. **c,** *In-situ* XRD $\theta$-$2\theta$ scans of SrCoO$_{3-\delta}$ recorded at 100 and 30 °C during cooling. The asterisk indicates x-ray radiation caused by the tungsten filament.